# Analytic Elastic Cross Sections for Electron-Atom Scattering from Generalized Fano Profiles of Overlapping Low-Energy Shape Resonances

Peter Nicoletopoulos[1]

Faculté des Sciences, Université Libre de Bruxelles, Brussels, Belgium

E-mail: pnicolet@skynet.be  and  pnicolet@ulb.ac.be

**Abstract**

The variation with energy of the total cross section, $\sigma_T(x)$, for elastic electron scattering from atoms of several elements is caused primarily by shape resonances corresponding to the formation of temporary negative ions. It is shown that such cross sections are expressible analytically in terms of a constant background B, added to a function F(x) given by a "generalized Fano profile" [Durand Ph, *et al* (2001) *J. Phys. B: At. Mol. Opt. Phys.* 34, 1953, *ibid* (2002) 35, 469]. In three cases (sodium, magnesium and mercury), a detailed consideration proves that the representation $\sigma_T(x)=F(x)+B$ is accurate in a fairly wide energy range. Moreover, the related momentum transfer cross sections, written in the form $\sigma_M(x)=aF(x)+B'$, are tailor-made for studying "elastic" electron transport in terms of the two-term solution of the Boltzmann equation: Not only are the resulting swarm transport coefficients adjustable to the experimental values (with $a \cong 1$, $B'<B$), but above all they are calculable very easily because the unnormalized energy distribution is obtainable analytically. The ample saving in computational effort is exploited in order to test the Wannier-Robson "momentum-transfer approximation"; it is found that, in these cases of resonance-dominated cross sections, the latter method can be so accurate that it can be used to recover $\sigma_M(x)$ from experimental data *algebraically*. In the margin, a model profile is presented that gives rise to negative differential conductivity.

-----------------------------------------------------------------------------------------------------

[1]Address for correspondence: Rue Joseph Cuylits 16, 1180 Brussels, Belgium.

# 1. Introduction

Unstable compound states (negative ion shape resonances) can be formed when an impacting low-energy electron is temporarily trapped in a neutral atom by a combination of polarization and centrifugal barrier potentials. The alkali metals, alkaline earths, and the group IB and IIB elements are particularly good candidates for such resonant behavior. As a result, the total elastic cross-sections, $\sigma_T$, for the scattering of electrons by atoms of these elements display large peaks at low energy. The asymmetric nature of the peaks suggests that these cross sections should be expressible in terms of the well-known Fano lineshape for an isolated resonance in a "flat" continuum (decay channel) [1]. However, a comparison with experimental cross sections shows that this is impossible: interference effects caused by higher-lying broad resonances imbedded in the same continuum significantly modify the single resonance Fano profile (a forceful example appears in section 6.1).

Lately, a simple and elegant theory has been presented in which the case of overlapping resonances is treated in terms of "generalized Fano profiles" $F(x)$ [2-3]. The addition of a constant background to these profiles provides analytic expressions for the elastic cross sections of group I and II elements that are quite precise in a wide energy range. This is the main theme of the present paper.

The incentive for this undertaking stems from the long-standing need to simplify the calculation of electron transport coefficients. In fact this work evolved in the course of an effort to model realistically certain transport theoretical phenomena in a classic variant of the Franck-Hertz experiment where the low-energy shape resonance in mercury plays an important role, as described in [4-5].

When inelastic collisions are inoperative, the energy distribution $f(x)$ of electrons traversing a gas is governed by the momentum-transfer cross section $\sigma_M(x)$. Of course, even in the simplest case of the two-term solution of the Boltzmann equation in the swarm approximation, the calculation of $f(x)$ is an unpleasant task because the cross section appears inside an indefinite integral. It was noticed that if $\sigma_M(x)$ is given by a conventional or a generalized Fano profile, this integral can be evaluated in closed form, and hence $f(x)$ and the resulting transport coefficients are obtainable with trifling computational discomfort.

Naturally, one is tempted to assume that the resonance profile of $\sigma_M(x)$ is similar to the expression derived for $\sigma_T(x)$ and inquire whether this form could be used to reproduce known experimental results. It will be shown that this is indeed accomplishable, with a different background B' and some minor tuning in the weight of F(x), namely: $\sigma_M(x)=aF(x)+B'$, $a\cong1$.

On the way to obtaining a good fit to experimental data it is found that some earlier determinations of momentum-transfer cross sections (biased as they often are by theoretical calculations) are faulty, in fact too narrow. What is lacking is the structure between the peak and the inelastic threshold caused by the overlap of the broader higher-lying resonances. This broad structure is precisely what the theory of generalized profiles is addressed with.

The result of this analysis is that there is a much wider range of E/N (the ratio of electric field to gas number density) in which inelastic effects are essentially inoperative. The reason is that, when driven through a gas of resonance-dominated atoms, a spatially homogeneous swarm acquires the character of a beam and essentially maintains this identity as the field rises, until the energy of its centroid attains the inelastic threshold.

As an aside, the present results provide a convenient means for testing the so-called "momentum-transfer approximation" initiated by Wannier [6] and developed further later on, especially by Robson [7-9]. This scheme is mainly useful for gaining a physical understanding of transport phenomena and is exact for constant collision frequency. It is found that in the present case of real gases with sharply peaked elastic cross sections the simple equations of that method can be very accurate. In fact, they can even be used for obtaining a remarkably good estimate of $\sigma_M(x)$ from experimental data on the drift velocity, without solving the Boltzmann equation.

The presentation is organized as follows: In section 2 we summarize the present knowledge on low-energy electron-atom shape resonances. In section 3 we briefly state the result of the theory of generalized Fano profiles and explain how these expressions can be brought into harmony with elastic cross sections of elements of section 2. In sections 4-6, we select three cases,

(Na, Mg, Hg, respectively) where sufficient information of resonance energies and widths is available, and construct generalized Fano cross sections consistent with existing experimental and theoretical results. For sodium and mercury, we use these expressions in order to fit measured transport coefficients and to test the momentum-transfer approximation. Section 7 describes a special kind of generalized profile leading to negative differential conductivity. A sharp multi-peaked profile similar to the theoretical low-energy cross section of cesium is given in section 8. Section 9 highlights a shortcoming of the method that is significant in the region of ultra-low energies.

## 2. Present knowledge on low-energy shape resonances

Work on low-energy shape resonances carried out before 1994 has been reviewed in [10]. Some newer results will be referenced in this summary.

In the alkalis, the ground state of the negative ion ($ns^2$ $^1S$) is stable. The first excited state (ns np $^3P$) is found to be an unresolved resonance in Li, Na, and K, that gives rise in each case to a pronounced peak in the cross sections at less than 0.12 eV. As to the heavier elements, a recent relativistic calculation shows that in Rb the fine structure separation is still small but in Cs the splitting is sufficiently larger than the widths, so that three peaks of enormous height (18000-24000 $a_0^2$) are displayed in the total cross section below 0.015 eV [11]. In Fr, the sharpness of structure is reduced as the widths increase.

Among the IB elements only Cu has been studied [12] and appears to have a sharp (4s4p $^3P$) resonance at 0.3 eV and a broader (4s4p $^1P$) feature at 0.5 eV.

In the lighter alkaline earths Be and Mg, the low-lying peaks are attributed to the unresolved ($ns^2np$ $^2P$) ground state of the negative ion that is now unstable. They are followed by a broad hump due to the $^2D$ state. In Ca, the ground state is stable but the $^2D$ resonance is narrower and gives rise to a well-defined peak in $\sigma_T(x)$ at about 1.1 eV. The result is that there is now a deep Ramsauer minimum behind the peak [13-14]. Analogous results hold for Sr. Little is known about Ba and next to nothing on Ra.

The group IIB elements are more accessible experimentally because the peak in the cross section produced by the lowest-lying ($ns^2np$ $^2P$) resonance occurs at several tenths of eV. As a result the existence of a "large variation in the mean free path" at about 0.5 eV in elastic electron scattering from Cd and Hg was discovered as early as 80 years ago [15] in a medium-pressure transmission experiment of the type discussed in [4]. The currently known experimental values for the energies of the peaks are 0.49 eV for Zn [16], 0.33 for Cd, [16,18] and 0.42 eV for Hg [17].

Theoretical work on Zinc and Cadmium is scarce. Values for the unresolved $^2P$ resonance energy of 0.23 and 0.31 eV have been obtained for Zn, and of 0.28 and 0.18 eV for Cd [19,20]. Much more is known about mercury. In this case there is substantial fine structure splitting of the doublet. Two peaks at energies of about 0.2 and 0.4 eV have been found for the $^2P_{1/2}$ and $^2P_{3/2}$ resonances in relativistic calculations but only a single peak is observed experimentally [17]. This discordance is resolved in section 6 when the profile of the peak is expressed in terms of overlapping resonances.

Low energy negative ion resonances in other groups have also been predicted sporadically (C, N, Pb, Tl).

## 3. Generalized Fano Profiles

The familiar formula due to Fano [1] for the asymmetric profile of the cross section in the presence of a single resonance interacting with a continuum (decay channel) has the form

$$\sigma = \sigma_a \frac{(q+\varepsilon)^2}{1+\varepsilon^2} + \sigma_b \qquad (1)$$

Here $\sigma_a$ and $\sigma_b$ are the resonant and nonresonant portions of the cross section, $\varepsilon = 2(x-x_0)/\Gamma$, with $x_0$ the resonance energy and $\Gamma$ its width, and q is a parameter called the index of asymmetry. In the case of one or more well-isolated resonances, equation (1) can be used to obtain q and hence the resonance parameters $x_0$ and $\Gamma$ from experimental profiles. In practice however, the complete profile is often the result of interference among overlapping resonances and the simple representation of the profile as a series of Fano terms is lost [21-22].

Recently, a generalized expression of cross section profiles involving two or more interfering resonances was derived in [2-3]. This lineshape is simply a product of Fano factors and of a Breit-Wigner term representing the portion of the continuum (lumped essentially into a broad "quasi-bound state", a wave packet) that is strongly coupled to the resonances. For two resonances, it has the form

$$\sigma(x) = \sigma_0 \frac{(q_1+\varepsilon_1)^2}{1+\varepsilon_1^2} \frac{(q_2+\varepsilon_2)^2}{1+\varepsilon_2^2} \frac{1}{1+\varepsilon_3^2} \qquad (2)$$

where $\varepsilon_i = 2(x-x_i)/\Gamma_i$, as before, and $\sigma_0$ is a constant. In this theory, the q's are not square roots of ratios of transition amplitudes as in the conventional Fano formula. They are given in terms of the roots $e_i$ of a certain polynomial $P(x)$ (of degree n for n resonances and one continuum), by $q_i = (2x_i/\Gamma_i) - e_i$. For n=1 and 2, the coefficients in $P(x)$ are simple functions of the complex energies $x_i - i\Gamma_i/2$ and certain "generalized" oscillator strengths. Such details are not required here but the point is that the expression of the profile in terms of q's is entirely a matter of convenience. A crucial feature of the form (2) is that, unlike (1), it tends to zero as x tends to infinity.

We will suppose that in the group I and II elements the cross section between zero and the higher-lying set of (Feshbach) resonances and Wigner-cusps associated with excited states can be written as

$$\sigma(x) = F(x) + B \qquad (3)$$

where $F(x)$ has the form (2) with a few shape resonances and one effective continuum, and B is a phenomenological quantity representing dynamics involving states that are not included in the model subspace spanned by the resonances, on which $F(x)$ is based. We will also assume that there is an energy interval $x_{min} < x < x_{max}$ in which B can be approximated by a constant. For the sake of simplicity we will take $x_{min} = 0$. In the cases considered here, $F(0) \cong 0$, so that $B \cong \sigma(0)$; if $x_{max}$ is sufficiently large, B is also close to $\sigma(x_{max})$.

Equation (3) will be used in the following sections in order to construct the elastic cross sections of Na, Mg and Hg. As we will see, these expressions are quite accurate in a region $x_{min} < x < x_{max}$ where $x_{max}$ can approach or even exceed the inelastic threshold and $x_{min}$ is of the order of thermal energies. Below $x_{min}$, they can be highly inaccurate (see section 9).

In building the profiles we proceed as follows. For our basic input we take the zeros (say $\alpha$ and $\beta$, $\alpha > \beta$) of the derivative of $\sigma(x)$. These are quantities that stand out in theoretical calculations and can be measured in an ideal beam experiment. In the absence of the damping factor the resonance parameters would be given by the simple expressions

$$x_1 = \frac{\beta + \alpha q^2}{1+q^2} \qquad \Gamma_1 = \frac{2q(\alpha - \beta)}{1+q^2} \qquad (4)$$

We begin by varying q until $x_1$ and $\Gamma_1$ are close to the best known values (a common property of these shape resonances is that $x_1 \cong \Gamma_1$ and $\beta \cong 0$; equations (4) then imply that $q \cong 2$). In the next step we insert these preliminary values in the full expression (3) with two factors in F(x) (one q), in order to display a first view of the generalized profile; the initial trial values of the "continuum" parameters $x_2$ and $\Gamma_2$ are determined by an educated guess with an eye on reproducing the observed form of the cross section between the resonance and the excitation threshold. Approximate values of the constants B and $\sigma_0$ are deduced from theoretical information on $\sigma_T(0)$ and on the height of the peak. In the end, the final values of q, $x_2$ $\Gamma_2$, $\sigma_0$, and B, are sought by adjusting to existing *experimental* data on $\sigma_T(x)$ without varying $x_1$, and $\Gamma_1$. The best fit might necessitate the inclusion of an additional asymmetric factor with a second profile index $q_2$.

Bare numbers for energy will refer to eV, for cross sections to squared Bohr radii ($a_0^2$) and for the ratio E/N to Td (1 Td=$10^{-17}$ V cm$^2$).

## 4. Sodium

*4.1. Derivation of the analytic cross section*

It seems that there is only one reliable measurement of $\sigma_T(x)$ for sodium, that of Kasdan *et al.* [23]. Their results agree well with the theoretical calculations of [24-26] and show that the older data of Brode [27] and Perel *et al* [28] are faulty. The recording of Johnston and Burrow [17] is distorted by the beam energy distribution and is thus unusable for our purposes.

A reasonable choice of input for sodium consistent with theory [25-26] and experiment [17] is given by $\alpha$ between 0.10 and 0.12 and $\beta$=0.01. For $\alpha$=0.108 the value q=1.73 gives $x_1$=0.083 and $\Gamma_1$=0.085, very close to the parameters found in [26] for the (3s3p $^3$P) resonance. Starting values for the remaining parameters can be obtained as follows: The computed cross sections of [25,26] suggest that $B \cong 300$ and $\sigma_0 \cong 620$ [since $\sigma_0 \cong (\sigma_{max}-B)/(1+q^2)$ and $\sigma_{max} \cong 2800$]. Further, one may suppose that the (3s3p $^1$P) state is a very broad resonance that is unobservable except insofar as it determines the parameters of the damping factor. In this spirit one can start with $x_2=0.15$ (as indicated by the early behavior of the $^1$P phase shift in [26]), and $\Gamma_2 \cong 1$.

Our final generalized Fano cross section fitted to the experimental data of [23] is shown in figure 1 (black curve). It is given by equation (3) with

$$x_1=0.083,\ \Gamma_1=0.085,\ q=1.3,\ x_2=0.11,\ \Gamma_2=1.5,\ \sigma_0=950,\ B=300 \qquad (5)$$

The ultimate value of $\alpha$ (the location of the peak) is 0.116, not far from our initial input and very close to the theoretical values in [24] and [25]. The lowest extremum $\beta$ has moved to 0.03.

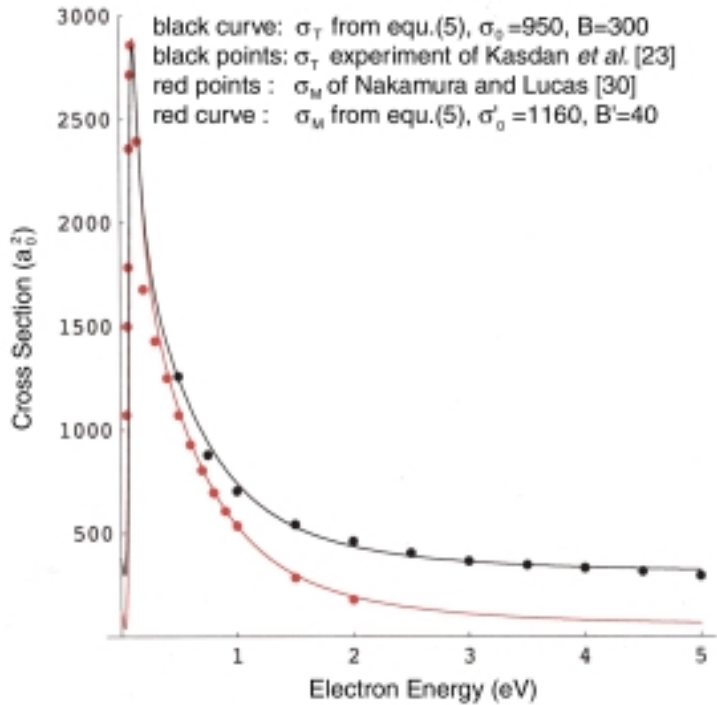

Figure 1

Although the peak itself is not included in the experimental points, their agreement with our cross section is very satisfactory. Obviously in this case the upper limit ($x_{max}$) of validity of the approximation "B=constant" extends well beyond the inelastic threshold (2.1 for Na). It should be stressed that the damping factor is essential for molding the energy behavior of $\sigma_T(x)$ between peak and threshold to the experimental curve, a point illustrated repeatedly in the next sections.

### 4.2. Calculating transport coefficients

We now assume that, for practical purposes, the inherent difference between $\sigma_M(x)$ and $\sigma_T(x)$ is essentially contained in the background term, so that $\sigma_M(x)=aF(x)+B'$, with $a\cong 1$, and use this expression to compute the drift velocity, as a function of E/N, of an electron swarm in sodium vapor, in the range of E/N where inelastic effects are negligible. The fact that $\sigma_M(0)$ differs from $\sigma_T(0)$ is not important since neither of these cross sections is expected to be accurate near x=0 (see section 9).

In the standard two-term approximation, the isotropic component $f_0(x)$ of the electron energy distribution is given by

$$f_0(x) = c\sqrt{x}\,\exp[-g(x)] \tag{6}$$

$$g(x) = \int_0^x \frac{dy}{kT + \frac{(eE/N)^2}{6(M/m)y[\sigma(y)]^2}} \tag{7}$$

$$\int_0^\infty f_0(x)\,d(x) = 1 \tag{8}$$

where e is the electron charge, k is the Boltzmann constant, T is the temperature of the neutral gas, E is the electric field, N is the number density of atoms and M/m is the atom/electron mass ratio ($205.5^2$ for Na).

With $\sigma(x)$ expressed in the form (3), the integral in equation (7) can easily be evaluated analytically, regardless of the number of factors in F(x). Therefore only the normalization factor, c, needs to be calculated numerically with (8) in order to obtain $f_0(x)$ and hence the anisotropic component $f_1(x)$ along the axis of the field. As a result, the various transport coefficients as functions of E/N are sequences of simple numerical integrals, computable in seconds. Our concern will be with the drift velocity w

$$w(E/N) = -\frac{1}{3}\frac{E}{N}\left(\frac{2e}{m}\right)^{1/2} \int_0^\infty \frac{x}{\sigma(x)} \frac{\partial f_0(x)}{\partial x} dx \quad (9)$$

and the mean energy $\varepsilon_m$

$$\varepsilon_m(E/N) = \int_0^\infty x f_0(x) dx \quad (10)$$

Strictly speaking, the upper limit of integration in equations (8-10) should be consistent with the limit of validity, $x_{max}$, of our expression for $\sigma_M(x)$. Fortunately, with the cross sections found here, the function $f_0(x)$ is essentially negligible beyond a low value of x, in a substantial interval of E/N. Therefore, for simplicity, all integrations were extended to infinity. For sodium, it was verified that even at the upper end of the interesting range of E/N, the results differed by only 2% from those obtained by cutting off the integrals at x=5.

The drift velocity w(E/N) of electrons swarming through sodium vapor was measured in [29] and was then used in [30] in order to recover the cross section. Unfortunately, these results must be regarded with suspicion. The data obtained are not independent of pressure: at a fixed E/N, the measured drift velocity *falls* at increased pressure. This was attributed to the presence of dimers (Na$_2$) and "true" drift velocities were obtained at each E/N by extrapolating the w(E/N,N) curves to zero pressure. However, as pointed out by Elford [36], the dimer effect should actually produce w(E/N,N) curves that *rise* with N. On the other hand, the observed behavior is consistent with spurious longitudinal diffusion due to deviations from spatial inhomogeneity, and indeed the latter effect might be unusually large because the drift space used was unduly short. But diffusion errors are known to *increase* at low pressure [36]. In other words, the w(E/N) curve obtained by extrapolating to zero pressure could well be *less reliable* than the higher-pressure curves.

It seemed advisable therefore to test our analytic cross section against *both* kinds of drift velocity data. The results are shown in figure 2. The red points are from the extrapolated data (assuming T=803° K) and the black points from uncorrected data at 803°. The full curves are obtained using (9) with $\sigma_M(x)=aF(x)+B'$, [equivalently $F(x)+B'$ with $\sigma_0$ changed to $\sigma'_0=a\sigma_0$] with the constants

$\sigma'_0 = 880$  (a≅0.93)  $B' = 200$  (red curve)
$\sigma'_0 = 1000$  (a=1.05)  $B' = 300$  (black curve)

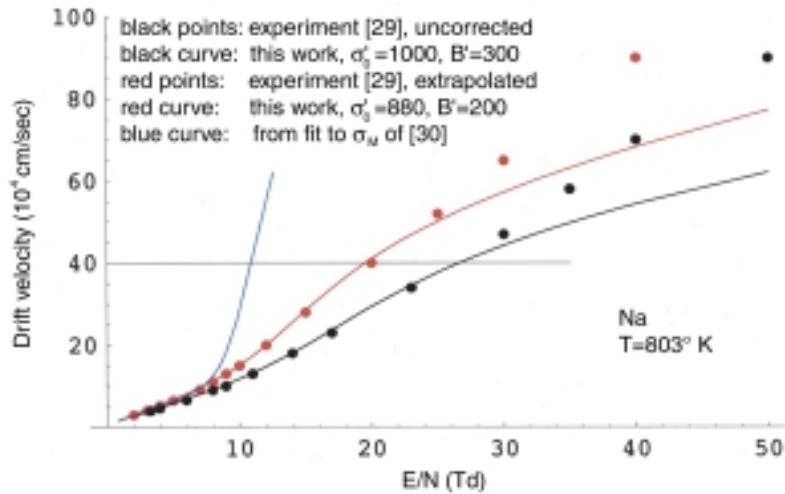

Figure 2

Both fits are satisfactory and show that the present method for obtaining realistic momentum-transfer cross sections is a good one. Needless to say, a more reliable set of drift velocity data would be required for establishing the analytic form of the true momentum-transfer cross section of sodium.

Regardless of this, the results of [29] are useful and raise some interesting points. To begin with, it seems that the red curve is essentially a redrawing of the black one on a different scale of E/N. This suggests that the difference between the two sets of data might be due, more or less, to a systematic error in the determination of N. If this is the case then all one has to do in order to obtain the true cross section is to find the correct value of B'.

Now let $E^*$ denote the value of E/N at which the transport coefficients begin to change appreciably due to inelastic impacts (particle-conserving effects–such as excitation and superelastic scattering–or "reactive" processes–such as cumulative ionization via metastable states, direct ionization, and attachment).

The trends, with respect to the experimental points, of the two "elastic" drift velocity curves in figure 2, point to values of about 20 and 30 Td for $E^*$.

These are respectively two and three times larger than the value obtained in [30], as deduced from the knee at E/N≅9 in their mean energy curve.

This disagreement is most likely a consequence of the well-known lack of uniqueness encountered in recovering $\sigma_M(x)$ from the data when inelastic collisions are incorporated in the Boltzmann equation. When that procedure was used in [30] (with the *red* points of figure 2) it produced a cross section that decreases very rapidly (see the red points in figure 1). A steeper-falling $\sigma_M(x)$ produces a longer-tailed energy distribution at low E/N, and hence a more significant fraction, R, of electrons with energies exceeding 2.1 eV.

Remarkably, that narrower cross section of [30] too, can be obtained from the basic profile (5) mainly by changing the value of B', namely with B'=40 and $\sigma'_0$=1160 (a=1.2), as shown in figure 1.

The value of the fraction R resulting from this form for $\sigma_M(x)$ at E/N=E*=9 is 50%. With our large-B' cross sections (respectively B'=200 and 300), R is 2.8% and 0.05% at 9 Td. But at E*=20 and 30 Td, R is indeed close to 50% (respectively 42% and 50%). Note that 50% need not be an implausibly large proportion because the ratio of the largest inelastic cross section to $\sigma_M(x)$ for x≅2.1 is less than 0.1.

What is interesting about these three numbers for E* is the corresponding value of the mean energy. The black upper and lower curves of figure 3 show the function $\varepsilon_m$(E/N) computed from the energy distribution resulting from the cross sections with B' = 200 and 300, respectively. We see that, at E/N=19.6 and 27, close to the approximate points where the drift velocity curves deviate from the respective experimental data, $\varepsilon_m$ actually reaches the inelastic threshold (say $\varepsilon^*$). Moreover, the critical drift velocity w(E*) is the same in both cases (about 40x10$^4$ cm/sec).

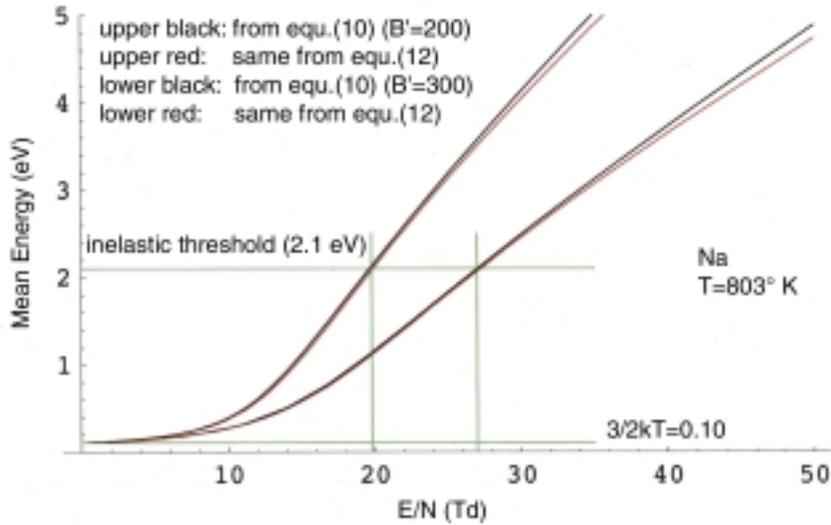

Figure 3

In contrast, the curve for $\varepsilon_m$ obtained in [30] barely attains 0.5 eV before tapering off at about E/N=9 Td; a similar limitation of the mean energy was found in [31], from the black data in figure 2, using the theoretical $\sigma_M(x)$ of [24] which is still narrower.

When w(E/N) is computed with our fit to the cross section of [30] (with B'=40), the deviation from the red points of figure 2, does in fact occur at about E/N=9 Td (see the blue curve in figure 2), and the corresponding mean energy is indeed only about 0.5 eV. However, w(E/N) slopes sharply upwards, so that the deviation from the experimental curve is now in the *opposite* direction. That type of behavior would require very large superelastic scattering and is therefore inconsistent with merely exciting collisions. Nor is it consistent with a simultaneous *decrease* in the mean energy (see subsection 4.3).

These objections indicate that the inversion procedures used in [30] and [31] biased as they are by the initial choice of too narrow an initial $\sigma_M(x)$, and perhaps on questionably accurate inelastic cross sections, lead to unphysical solutions.

This should be avoidable by starting the inversion procedure with our Fano-type cross-section since the correct direction of the bend at E* imposes a lower limit $B_{min}$ on B'. The value of $B_{min}$ can be found by trial and error. For the red data of figure 2 it is indeed about 200.

We should note that there is also an upper limit $B_{max}$. This can be seen by noticing that the function $\sigma_M(x)$ leading to the black curve of figure 2 is very close to $\sigma_T(x)$ since B'=B. But in adding a constant B to F(x) in (3) in order to fit the total cross section, we are flirting with the unitary limit. Since a≅1, it follows that a value of B' larger than B is liable to violate unitarity and so we can say that $B'_{max}$≅300. Thus in the true $\sigma_M(x)$ we should have 200<B'<300.

These remarks uphold the appropriateness of the present method in obtaining momentum-transfer cross sections.

The essence, however, of this argumentation is that in sodium (and by analogy in other elements of section 2), the elastic cross section seems to impart a beam-like character to the swarm so that

(a) E* is the value of E/N where $\varepsilon_m(E/N) \equiv \varepsilon^*$.

(b) E* is also obtainable from the onset of deviation of the purely "elastic" drift velocity w(E/N) from the experimental data.

And incidentally

(c) A fortuitous systematic rescaling of E/N will not affect the values of w(E*) and $\varepsilon_m(E^*)$.

Notice that in the neighborhood of the inelastic threshold, $\sigma_M(x) \cong B'$. Statement (c) then indicates that $\varepsilon_m(E^*)$ and w(E*) are both proportional to $E^*/\sigma_M(\varepsilon^*)$, so that if $B'_1/B'_2 \cong 2/3$, then $E^*_1/E^*_2 \cong 2/3$, as in figure 2.

A more elaborate computational effort incorporating inelastic cross sections is required in order to ascertain these inferences. But their plausibleness can be demonstrated in terms of a simple physical framework, as will be seen in the next subsection.

*4.3. Testing the momentum-transfer approximation*

As mentioned in section 1, a simple means for discussing transport phenomena without solving the Boltzmann equation is provided by the so-called "momentum-transfer approximation". It is interesting to see how our Boltzmann results compare with those obtainable from the basic equations of that approach [7-9].

For a spatially uniform swarm without a magnetic field, in the center of mass of the atoms and the colliding swarm particles, these equations are

$$w = \frac{e}{(2m)^{1/2}} \frac{E}{N} \frac{1}{(\varepsilon_m)^{1/2} \sigma_M(\varepsilon_m)} \tag{11}$$

$$\varepsilon_m = \frac{3}{2}kT + \frac{1}{2}Mw^2 - \Omega(\varepsilon_m) \tag{12}$$

Notice that in (12), M is the atomic mass. The term $Mw^2/2$ represents that part of the energy that has been provided by the electric field but randomized by elastic collisions.

The function $\Omega$ describes the energy loss due to inelastic or reactive effects. In the present case of homogeneous swarms it is *positive*, (excepting the unlikely case of very strong superelastic collisions). Its exact form (see [7-8]) is not required here; the important point for our purposes is the behavior of this function below threshold (namely $\varepsilon_m < \varepsilon^*$). Model forms of $\Omega(\varepsilon_m)$ based on a Maxwellian energy distribution extend below threshold (see [7]) but it is plausible that with the narrow distributions induced by the cross sections of certain elements, this tail becomes negligible so that $\Omega(\varepsilon_m) \cong 0$ for $\varepsilon_m < \varepsilon^*$.

At $\varepsilon_m \cong \varepsilon^*$, we can set kT=0 for simplicity and solve (11) and (12) for w and $\varepsilon_m$ near $\Omega=0$, (where $\varepsilon_m = \varepsilon^*$, E/N=E*, $\sigma_M(\varepsilon_m) = \sigma^*$). The result is

$$\varepsilon_m = a_1(E^*/\sigma^*) - a_2 \Omega$$
$$w = a_3(E^*/\sigma^*) + a_4 \Omega$$

where the quantities $a_i$ are constants.

These equations explicitly incorporate the statements (a), (b) and (c) of the previous subsection. Moreover, they predict the correct directions for the bends away from the data in the $\varepsilon_m(E/N)$ and $w(E/N)$ curves at $E/N \cong E^*$.

An exact calculation of the function $\Omega(\varepsilon_m)$ requires a good knowledge of several inelastic cross sections and is hardly an effortless task since the latter are not expressible analytically. Therefore it was not attempted. Hence, the validity of assertions (a-c) for sodium remains inferential.

It is easy however to test equations (11-12) in the "elastic" regime. The red curves of figure 3 shows the mean energy computed from equation (12), with $\Omega=0$, using our interpolations of figure 2 to the two sets of experimental drift velocity data. The agreement between these curves and the exact ones in the range $0<E/N<30$ is very good and shows how precise equation (12) can be for calculating the mean energy in terms of the drift velocity data.

We can now calculate the function $\sigma_M(\varepsilon_m)$. Equations (11-12) imply that

$$\sigma_M(\varepsilon_m) = \frac{M}{m} \frac{E}{N} \frac{1}{[2\varepsilon_m(\varepsilon_m - 3kT)]^{1/2}} \qquad (13)$$

where of course $\varepsilon_m = \varepsilon_m(E/N)$ is given by one of the red curves of figure 3 in their respective ranges $E/N < E^*$.

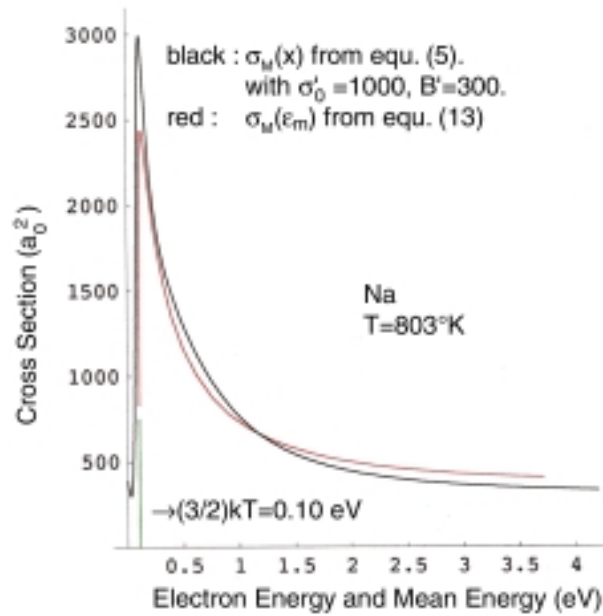

Figure 4

Figure 4 shows the "cross section" $\sigma_M(\varepsilon_m)$ computed in this manner, in terms of the lower curve of figure 3 for $\varepsilon_m(E/N)$, along with the corresponding $\sigma_M(x)$ (the one with B'=300). Clearly, for x>3/2kT, there is essentially a one-to-one correspondence between x and $\varepsilon_m$. A similar agreement was obtained for the cross section with B'=200. Consequently, a good set of data on w(E/N) and the simple momentum-transfer approximation in terms of the *mean energy* of the swarm can be sufficient for obtaining a very good approximation for the profile of the momentum transfer cross section, as well as its absolute magnitude, as a function of *electron energy*. The exact location of the peak of $\sigma_M(\varepsilon_m)$ in figure 4 is at $\varepsilon_m$=0.124 (at which point E/N=1.96); the peak of $\sigma_M(x)$ is at x=0.116. Note also that as E/N tends to zero and $\varepsilon_m(E/N)$ approaches 3/2kT, the right-hand side of (13) converges; $\sigma_M$(3/2kT) was found to be about 400 .

The derivation of equation (11) relies on the assumption that the energy distribution is sharply peaked at $\varepsilon_m$, so that in averaging the collision frequency $\nu$ [where $\nu(x)=N(2x/m)^{1/2}\sigma_M(x)$] the main contribution comes from energies around $x \cong \varepsilon_m$ (see [7]). This premise is clearly borne out remarkably well in sodium, despite the fact that $\sigma_M(x)$ is not a slowly varying function of x in the region of the peak. Obviously, some rather miraculous fine-tuning is taking place in this case, and we should not expect to find the same degree of agreement in other elements.

But regardless of such details, the above considerations implicitly support our previous inference that, in the elements considered here, the "elastic" electron swarm acquires the identity of a blunt-fronted beam and maintains this character throughout a considerable interval of E/N.

How close that interval is to the point where the corresponding mean energy reaches the inelastic threshold can perhaps be ascertained by averaging the inelastic contributions in order to obtain the function $\Omega$.

The matter of condition (a) will be revisited and further elucidated in section 6.2 in connection to mercury.

## 5. Magnesium

Theoretical values of the resonance energy $x_1$ for Mg calculated in the literature [10] are between 0.14 and 0.16 eV. The values found for the width are very similar except for one case [32] where $\Gamma_1$=0.24. However, equation (4) shows that there is an upper bound $\Gamma_{max}$ for $\Gamma_1$. In particular, for $\beta=0$, $\Gamma_{max}=\alpha$, (at q=1) and so the values $\alpha=0.19$, $\Gamma_1=0.24$, obtained in [32] are incompatible with an initial estimate based on the Fano formula. Therefore we assumed that once more $x_1 \cong \Gamma_1$.

We base our fit on the premise that $\alpha=0.18$, $\beta=0.005$. For q=1.9, the resonance parameters resulting from (4) are $x_1=0.142$, $\Gamma_1=0.144$. Figure 5 shows our estimated Mg cross section obtained with (3) with the set

$$q=1.9,\ x_1=0.142,\ \Gamma_1=0.144,\ x_2=1.74,\ \Gamma_2=5,\ \sigma_0=800,\ B=71 \qquad (14)$$

where B is deduced from the calculation of the scattering length in [13]. With this value of B, the choice $\sigma_0=800$ places the peak height at about the value found in [32]. The final values of the extrema are $\alpha=0.182$, $\beta=0.005$.

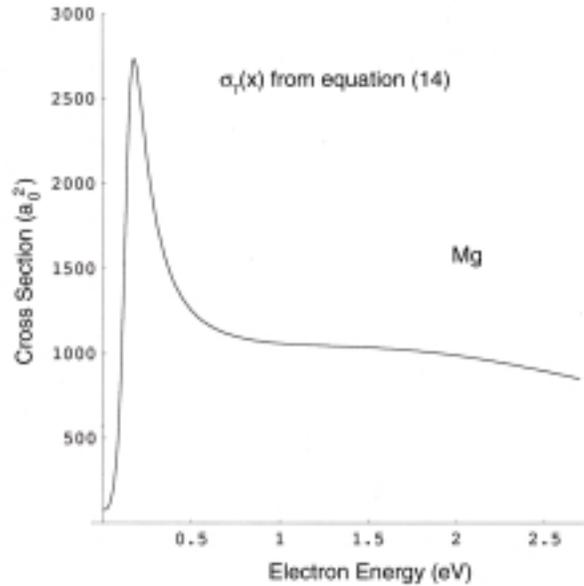

Figure 5

Unfortunately, no transport results are available for testing the analogous momentum-transfer cross section. The only experimental information available for comparison is contained indirectly in the transmission experiment of Burrow *et al.* [16]. A signal of the resonance in the negative derivative of the transmitted current was reported in [33] but this is too distorted by the beam energy distribution to be usable for a meaningful comparison.

In figure 6, the constant A has been adjusted in order to fit the expression exp[-AF(x)] to the experimental curve of [16] within the leeway permitted by the stated 0.03 eV uncertainty in the energy scale. In view of the lack of corrections for the beam width, the agreement is as good as can be expected. Of course, this result is independent of the values of $\sigma_0$ and B. It does indicate, however, that the hump attributed to the ($3s^23p\ ^2D$) state is reproduced correctly in our generalized Fano profile. A similar agreement was obtained in [32] in terms of their calculated cross section.

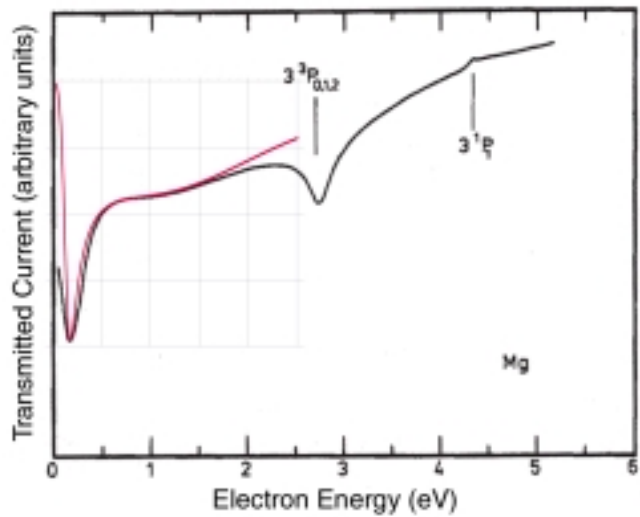

Figure 6

## 6. Mercury

### 6.1. Derivation of the analytic cross section

In comparison to other elements, the list of experiments dealing with the low-energy shape resonance in mercury is quite numerous. The initial determinations of the location of the peak in the cross section were obtained from swarm data in [34] and from a transmission experiment in [16]. These values were very similar (about 0.65 eV). Unfortunately they were both wrong. The former because the drift velocity data used was inaccurate due to $Hg_2$ formation and to diffusion effects [36]. The latter because of a fortuitous miscalibration of an X-Y recorder [41].

Subsequent beam experiments [35,17] placed the peak of $\sigma_T$ at about 0.4 and 0.42±0.3 eV, respectively. And a first attempt to obtain $\sigma_M$ from w(E/N) data (at T=573°, corrected for dimer and diffusion effects) showed a peak at about 0.5 eV [36]. Later on, the use of mixtures in order to weaken the pressure dependence of the dimer effect, and a thorough treatment of diffusion errors, lowered the swarm value of the peak to 0.44 [37]. Thus, given the different uncertainties in beam and swarm data, it seems safe to say that the "best" value for the peak in either cross section is close to 0.4 eV.

In regard to theory, the treatment of an element as heavy as mercury must include relativistic effects. Such a calculation was first reported in [38]. It was found that there is substantial fine structure splitting of the doublet lowest state of the negative ion, leading to two resonances at about 0.2 and 0.4 eV corresponding to the $^2P_{1/2}$ and $^2P_{3/2}$ states. The latter is broader than the former but apparently not broad enough to prevent the formation of two distinct peaks in $\sigma_T(x)$. A different theoretical treatment [39] also gave separated peaks at somewhat different energies.

The clear definition of peak separation displayed in [38-39] would doubtless show up in the transmission experiment of Johnston and Burrow [17]. Its absence has led to some confusion, as underlined in [10].

Initially, the author of [38] argued that the true width of the $^2P_{3/2}$ resonance (presumably the value obtainable by including a larger set of states in the decay channel) is such that this resonance will be indistinguishable from the background.

An alternative is to adopt the attitude that because the results are very sensitive to the precise form of the polarization potential, one must seek a parametric form for this potential that can be adjusted to give a peak in the cross section at the observed energy. This artifact is able to produce a momentum-transfer cross section that is not greatly dissimilar from the one deduced from swarm data in [37]. A different theory based on the same principle was presented recently in [40]; but while the peak in the resulting momentum-transfer cross-section is in excellent agreement with experiment, the total cross section is not: the peak in $\sigma_T$ appears at 0.55 eV, in clear conflict with beam data.

As we will see, the present work resolves this controversy. Namely the energies resulting from the *ab initio* theory of [38] are entirely consistent with a single-peaked generalized Fano profile formed by two overlapping resonances at about 0.2 and 0.4, with widths of 0.34 and 1.1, respectively.

Unlike the cases of Na and Mg, there is now a sufficiently undistorted experimental profile of the entire peak region (figure 1 of [17]) on which we can base our construction without unconvolving the beam energy distribution. This graph is proportional to the negative of the derivative of $\sigma_T(x)$ and gives the values of $\alpha$ and $\beta$ directly. Actually I used a copy of this recording that extends to a wider energy range [41].

The first step is to write the conventional Fano form (1) in terms of $\alpha$ and $\beta$ using equation (4), with $\alpha=0.42$ and $\beta=0.01$, and to vary q until the negative derivative of this expression matches the experimental curve. It turns out that this is feasible up to a positive additive constant. A good fit is obtained with q=0.6 by starting with $\alpha=0.51$ and adding about $(3/5)\sigma_0$ to the negative derivative. This combination drives $\alpha$ back to 0.42 and gives $x_1 \cong 0.14$, $\Gamma_1 \cong 0.44$. But the need for the additive constant in the derivative implies that equation (1) does not adequately represent the dynamics, since a steeply decreasing background contribution [roughly $-(3/5)\sigma_0 x$] is being left out of the integrated cross section.

In the next step we use these preliminary values in the (single q) resonance form of equation (2) and try to obtain a fit with reasonable values of $x_2$ and $\Gamma_2$. Satisfactory agreement is obtained with the set of parameters

$$q=0.61, \; x_1=0.145, \; \Gamma_1=0.39, \; x_2=0.34, \; \Gamma_2=1.65 \qquad (15)$$

where every effort was made to make $x_1$ as large as possible. The final value of $\beta$ is 0.025. Remarkably, the energy-dependent background found in the first step is now *automatically* absorbed into the Breit-Wigner factor. Yet this result is still incompatible with theory because $x_1$ is 30% lower than the calculated value.

In the final step we use equation (2) with two resonances. A cross section matching the experimental curve is now obtainable with a larger value of $x_1$. This fit is shown in figure 7. The parameters are

$$q_1=1.85, \; x_1=0.185, \; \Gamma_1=0.34, \; q_2=2.1, \; x_2=0.395, \; \Gamma_2=1.1, \; x_3=0.9, \; \Gamma_3=2.9 \quad (16)$$

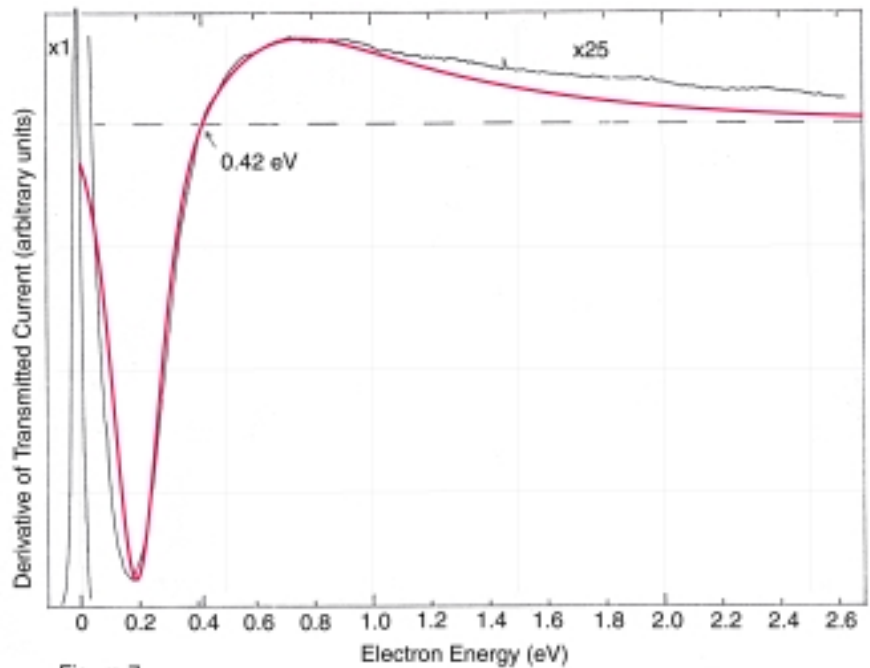

Figure 7

Both $x_1$ and $x_2$ are now close to the theoretical values of Walker [38]. The final value of β is negative. Apart from that, the profiles of steps 2 and 3 are almost completely identical. The mismatch at very low energy is inherent to the method and is not germane in the context of swarm experiments (see section 9).

These results show how fruitless it is to guess the energies (let alone the widths) of these shape resonances in terms of the peak in the cross section. In the cross section defined by (16), the peak is actually close to the energy of the $^2P_{3/2}$ state, whereas the $^2P_{1/2}$ state has the character of a "window" resonance since its position coincides with the lower extremum of the *derivative* of $\sigma_T(x)$.

The onset of disagreement with the experimental curve on the high-energy side indicates that our profile levels off too early. This can probably be corrected by incorporating more resonances (the $^2D$ states), but we have not gone that far in this work.

Figure 8 shows a fit to the total cross section of Jost and Ohnemus [35], obtained with our profile (16) and

$$\sigma_0 = 43.2 \qquad B = 300 \qquad (17)$$

We see that for 0.15<x<2.3, more or less, our analytic $\sigma_M(x)$ matches the cross sections of both beam experiments.

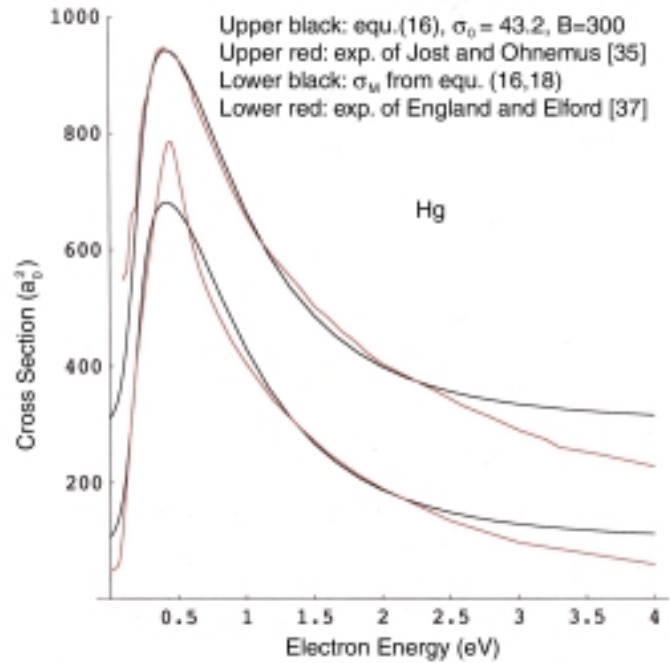

Figure 8

## 6.2. Calculating transport coefficients

As with sodium, we construct a momentum transfer cross section in order to compare the drift velocity w(E/N) (computed by equation (9) with T=573° and M/m=607²) to the experimental results. A large collection of the carefully corrected data of [37] is listed in [40]. Figure 9 shows the fit to these points obtained with $\sigma_M(x)=F(x)+B'$ using the (double q) profile defined by (16), and

$$\sigma'_0=39.3 \qquad\qquad B'=97 \qquad\qquad (18)$$

An identical fit is obtained from the (single q) profile of equation (15), with the constants

$$\sigma'_0=442.5 \qquad\qquad B'=91 \qquad\qquad (19)$$

Evidently, the swarm data in this range of E/N is insensitive to the small difference of these profiles at very low energy.

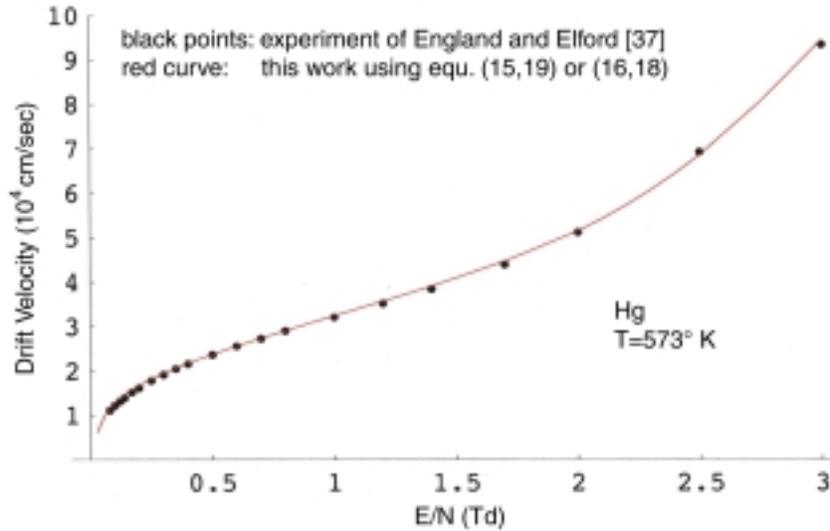

Figure 9

At first sight the agreement with experiment of our drift velocity in mercury is excellent. It must be pointed out, however, that to some extent this might be fortuitous. Unlike sodium, the resemblance of our $\sigma_T(x)$ to the experimental total cross section is not good all the way to threshold. The same is true for $\sigma_M(x)$, as seen in figure 8 where our momentum-transfer cross section is compared to the one derived in [37] from the same drift velocity data. Obviously, with the present profile $F(x)$, the upper limit, $x_{max}$, of validity of the approximation "B=constant" does not exceed 2.5 eV. As a result, the highest calculated drift velocities are uncertain. Indeed, at the upper end E/N=3 of the range of E/N, our values (obtained by integrating to infinity) differ by 8% from those obtained by cutting off the integrals at x=5. For sodium, one could calculate w(E/N) all the way to $E^*$ before that difference exceeded 2%.

Therefore, for mercury, any definitive claims in regard to $E^*$ and to inferences (a-c) of section 4.2 must await the refinement of our present profile by incorporating more resonances.

On the other hand, since there is no sign of departure of our computed w(E/N) curve from the experimental points, one might disregard these misgivings and still compute the value of $E^*$ from condition (a). The black curve in figure 10 shows the mean energy $\varepsilon_m(E/N)$ obtained by equation (10) using the energy distribution arising from either of our two cross sections. The value of $E^*$ resulting from this graph is about 5.5.

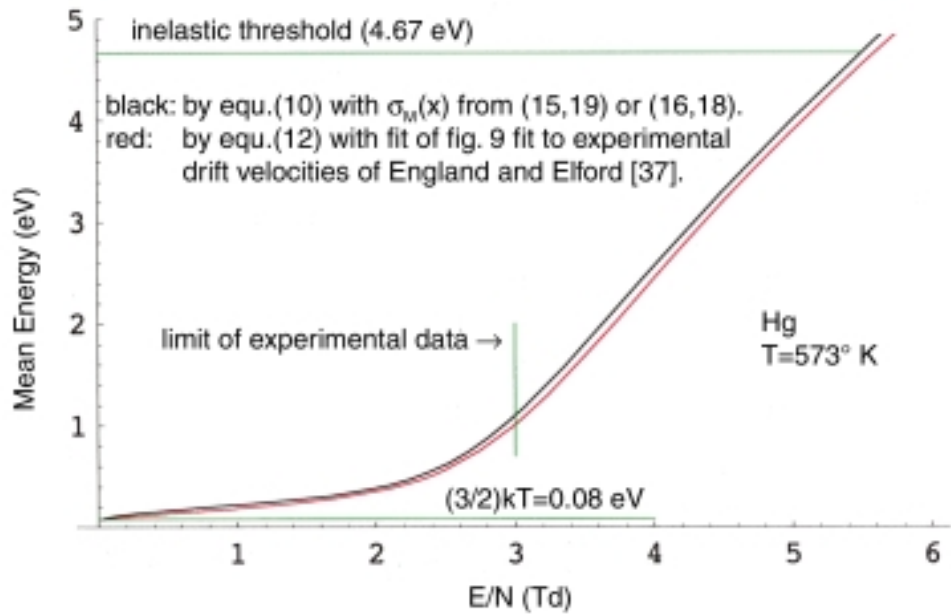

Figure 10

The fraction R of electrons with x>4.67 calculated with our energy distribution is less than $2 \times 10^{-11}$% at 1 Td and it is still only 2.8% at 3 Td. At 5.5 Td (our presumed value for E*) it reaches 43%, which was about the proportion needed for producing an observable deviation from the data of the elastic w(E/N) curve of sodium (see section 4.2).

It appears therefore that in mercury too, a scenario much like (a) and (b) of section 4.2 is probably taking place.

Collateral support is obtained by observing the evolution of the energy distribution (see figure 11). We see that, as E/N is raised, the increase in the fraction R is not due as much to a lengthening tail, as to the buildup of a second hump in the distribution. As E/N is raised, this hump moves toward higher energy and grows at the expense of the low energy peak.

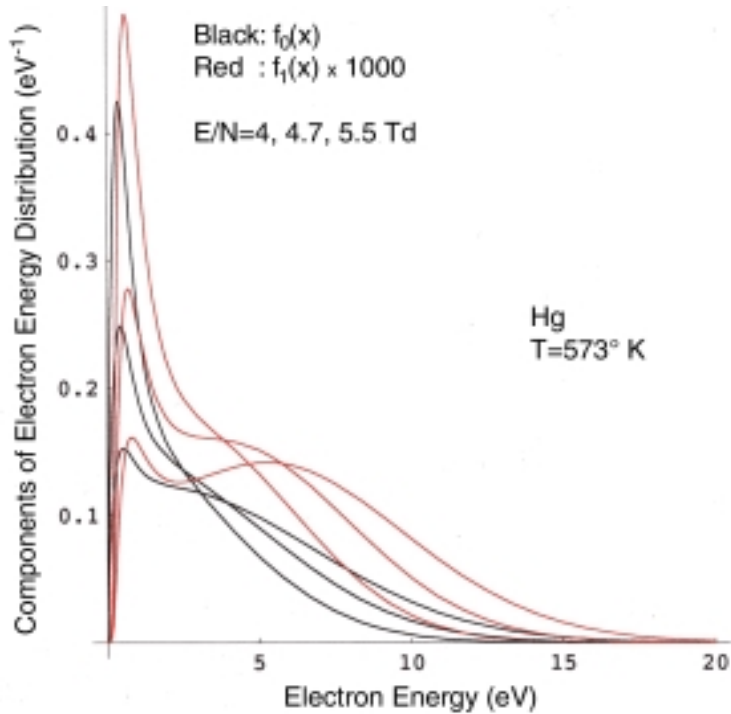

Figure 11

This is borne out experimentally, because the electron energy distribution in mercury has actually been measured, on two occasions. The first measurement was carried out a while ago in a "low-voltage" arc [42] and the second more recently [43] in a Franck-Hertz experiment. These are only piecemeal swarm experiments since, overall, the effect of spatial density gradients and space charge phenomena is far from negligible. Still, they both show the presence of two distinct peaks. In these measurements, however, the low-energy peak (at less than 1 eV as in figure 11) is substantially more pronounced, while the high-energy peak is weak and centered *above* the inelastic threshold.

Such evidence of clean-cut peak formation suggests that condition (a) is indeed implemented and might in fact be enforced *self-consistently.* Below the critical value of E/N, as the higher-energy hump grows, any small portion of its tail intruding past the excitation threshold is quickly clipped off by inelastic collisions, with resulting blunting of the front and replenishment of the low-energy peak. In this way, the swarm is even more likely to reach the inelastic threshold in bulk. At that point, the population of the higher-energy group declines sharply and the low-energy peak is enhanced dramatically, as indicated by experiment.

## 6.3. Testing the momentum-transfer approximation

As with sodium, it is worthwhile to compare the results obtained from the Boltzmann equation to those found from the momentum and energy balance equations of the momentum-transfer approximation.

The red curve of figure 10 shows the average energy $\varepsilon_m(E/N)$ calculated from equation (12) using our computed interpolation $w(E/N)$ to the experimental drift velocity in the range of available data. Figure 12 displays the function $\sigma_M(\varepsilon_m)$ obtained with these values of $\varepsilon_m(E/N)$, along with $\sigma_M(x)$.

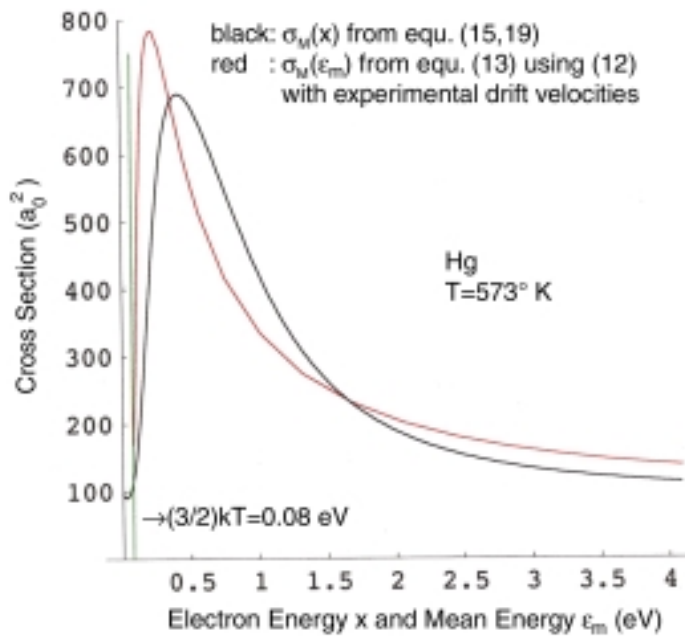

Figure 12

In both cases the agreement is less impressive than in sodium, but is still qualitatively correct. The mean energy runs parallel to the Boltzmann-equation curve but is somewhat lower, while the function $\sigma_M(\varepsilon_m)$ calculated from equation (13) is narrower than $\sigma_M(x)$ with a peak at 0.25 eV (at which point E/N=1.4). The lower limit $\sigma_M(3/2kT)$ of the function $\sigma_M(\varepsilon_m)$ was found to be 52 and 65, respectively, for the $\sigma_M(x)$ of equations (15,19) and (16,18).

Obviously, the degree of deviation of equation (12) from the true mean energy and the closeness of $\sigma_M(\varepsilon_m)$ to $\sigma_M(x)$ depend on the detailed structure of the peak profile. Expectedly, not all profiles conspire as well as they do in sodium for producing a one-to-one correspondence between x and $\varepsilon_m$.

## 7. A model for negative differential conductivity (NDC)

Figure 13(a) shows three hypothetical "mercury" momentum-transfer cross sections given by equations (2-3) with a single resonance, for different values of q, and corresponding ones for $\sigma_0$, namely

$$q=\{0.8; 1.2; 1.6\}, x_1=0.5, \Gamma_1=0.2, x_2=0.8, \Gamma_2=4, \sigma_0=\{600; 400; 280\}, B=10 \quad (20)$$

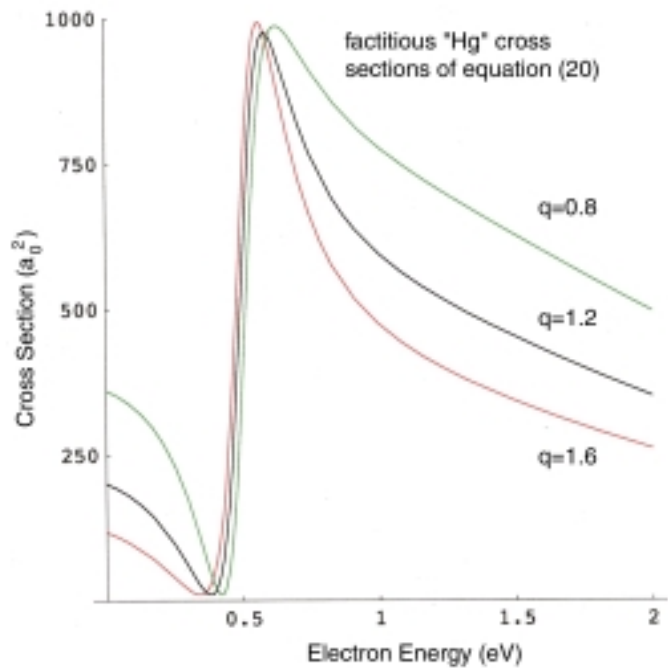

Figure 13(a)

They all display a deep Townsend minimum followed by a peak of about 1000 $a_0^2$ at x≅0.6 eV. Figure 13(b) shows the drift velocity resulting from these cross sections at T=573° with the mercury value of M/m. The black and red curves are very similar in shape and magnitude to the uncorrected experimental w(E/N) curves for mercury obtained originally in [44] and later in [30] and [36]. This explains why the early derivations of $\sigma_M(x)$ for Hg from inversion of uncorrected swarm data produce a peak at about 0.6 eV, preceded by a very precipitous cliff as in [34], and can even feature a substantial dip as in [30].

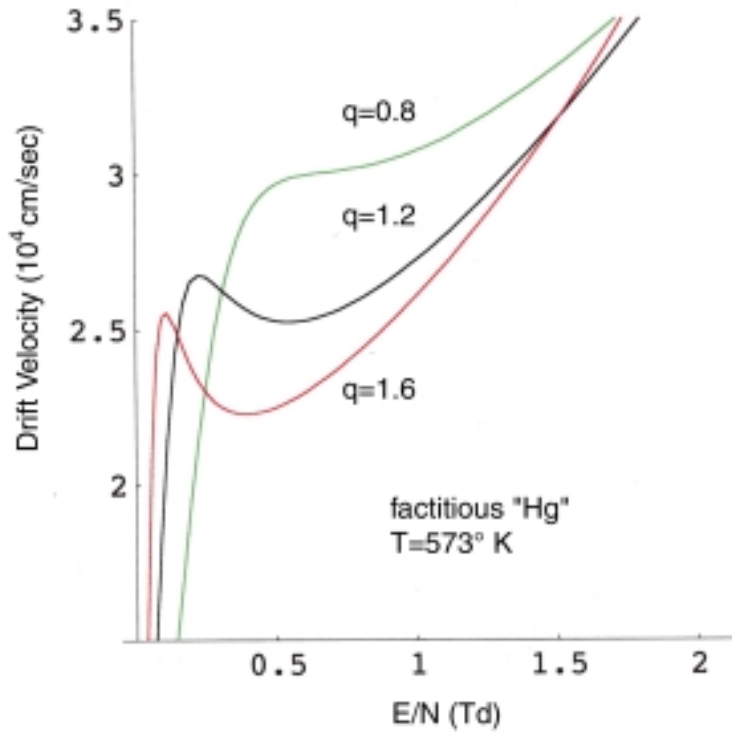

Figure 13(b)

It is currently accepted [36-37] that in mercury the change of slope of w(E/N), in a certain interval of E/N, leading to negative differential conductivity (namely dw/dE<0) is due to the presence of dimers. The above examples show that this effect can be modelled by an effective momentum-transfer cross section whose profile displays a substantial dip molded by a resonance. An additional constraint is that the resonance profile must be rather symmetric, namely q>1. The larger the value of q the deeper is the minimum of w(E/N). In our examples the minimum in w(E/N) disappears for q≅0.8, even though the dip in σ(x) is more pronounced [see figure 13(b)].

Transport theory has led to the belief that the NDC phenomenon can not occur for electrons in a pure gas in the presence of only elastic collisions. One needs either a *mixture* of monatomic gases [45] or some kind of inelastic process [7].

The latter case has been more fully illustrated recently by Blake and Robson [46] in terms of equations (11-12) using appropriate models of $\Omega(\varepsilon_m)$. It is interesting to note that the "shape" requirement $d\Omega/d\varepsilon_m<-1$ for NDC, established in [7] and [46] for inelastic collisions, is very similar to our asymmetry criterion q>1.

Obviously, a mixture of atoms and their dimers produces an effective cross section of this type. But perhaps NDC could arise from purely elastic scattering in elements such as Ca and Sr by means of the deep Ramsauer-Townsend minimum mentioned in section 2. However, as noted in section 9, this cannot be verified by constructing a generalized Fano profile unless the present prescription is amendable in order to include the contribution of the s-wave.

It is easily verified that for those cross sections of (19) where q>1, the momentum-transfer approximation fails completely in the critical region of E/N. After the initial rise, the mean energy derived from equation (12) does not merely level off close to the true mean energy, as in figure 10, but acquires a minimum. The result is that the function $\sigma(\varepsilon_m)$ satisfying the system of equations (11,12) is hardly a physical cross-section since it becomes double-valued in the region of the minimum. Understandably, the storage of the field energy is interrupted, and the term $Mw^2/2$ is meaningless in the critical region.

The model cross sections of figure 13(a) can also be used to illustrate the formation of double-peaked energy distributions that have been observed to accompany the onset of NDC. It is readily calculated that both $f_0(x)$ and the anisotropic component $f_1(x)$ display two peaks separated by a deep valley located at the energy where the cross sections go through a minimum. It should be noted however that this occurs even in the case where q=0.8 and is thus independent of NDC itself, as in the analogous case involving inelastic collisions [46]. Similar double peaked energy distributions were shown to arise in a swarm through molecular nitrogen and were interpreted in terms of vibrational excitation in collaboration with momentum transfer collisions [47].

## 8. A multi-peaked profile.

So far, we have dealt with overlapping resonances producing a single peak in the elastic cross section. One may wonder how the method of generalized Fano profiles fares in cases where the cross section displays several sharp peaks, and whether any of this finer structure shows up in the function $\sigma(\varepsilon_m)$ calculated from equation (13). To answer these questions we constructed a profile resembling the theoretical cross section of cesium computed in [11]. The result is shown in figure 14 (blue curve). The parameters (with energies in meV) are given by

$$q_1=15,\ x_1=2.5,\ \Gamma_1=1.1,\ q_2=1.9,\ x_2=6,\ \Gamma_2=1.15,\ x_3=15,\ \Gamma_3=6 \tag{21}$$

$$\sigma_0=4300 \qquad\qquad B=1000 \tag{22}$$

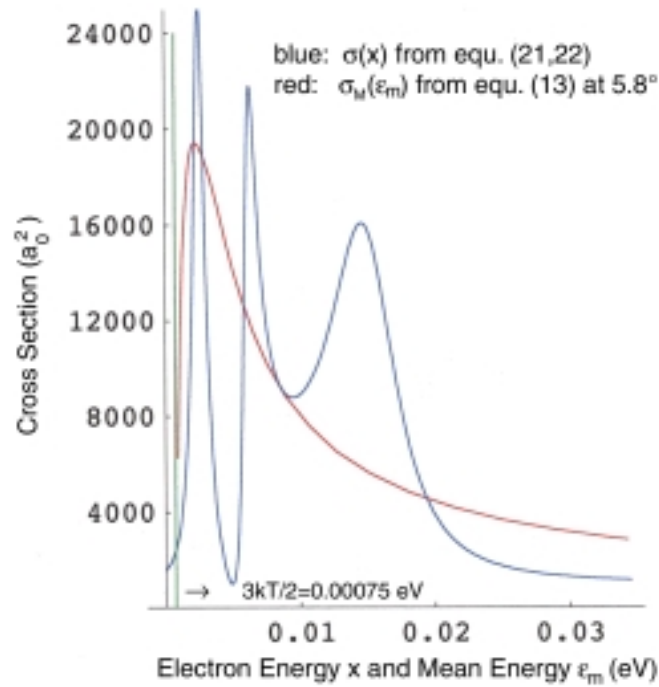

Figure 14

The red curve in figure 14 shows $\sigma_M(\varepsilon_m)$ computed at T=5.8° K. As we see, the structure is blurred into a single peak close to the first resonance. We note that the isotropic energy distribution is double peaked, but the anisotropic component contains three distinct peaks at low energy (for E/N≅0.04), as well as a fourth at higher energy. In both components, the extra higher-lying peak is analogous to the higher-energy hump in mercury mentioned in section 6.2.

## 9. The region of ultra-low energies

We conclude this paper with some comments on the shape of our generalized Fano cross sections in the region of thermal energies.

For mercury, a test of the accuracy of $\sigma_M(x)$ near x=0 is available [37,40] in terms of the thermal diffusion coefficient $ND_{th}$ at 470° K measured in [48]. This experimental value is $1.74\pm0.17 \times 10^{21}$ cm$^{-1}$ s$^{-1}$. The number obtained with our cross section of equations (16,18) is $1.05 \times 10^{21}$ cm$^{-1}$ s$^{-1}$, which is too low. The authors of [40] point out that an improved value of $ND_{th}$ is obtainable if the cross section displays a dip near zero energy. Such a dip occurs in the cross section defined by equations (15,19) and indeed this profile yields the better value $ND_{th}=1.32 \times 10^{21}$ cm$^{-1}$ s$^{-1}$. The actual presence of a minimum in the total cross section at about 0.1 eV is suggested from the experimental curve of [17] (see figure 7). In $\sigma_M(x)$, it might be more pronounced since B'<B and ultimately $\sigma_M(0)$ will have to be equal to $\sigma_T(0)$.

Having said this, we must emphasize that, as it stands, our approach is in principle unable to represent the cross section accurately at ultra-low energies. The formation of the lowest extremum of $\sigma_T(x)$ is often affected considerably by a locally rapidly varying s-wave contribution (see [13]). Therefore, neither is the true value of β related entirely to the effective Hamiltonian on which (2) is based, nor is it justifiable to assume that the quantity B is constant throughout the low-energy region.

Perhaps one can extend $\sigma_T(x)$ all the way to zero by adding a virtual (negative energy) state. But this form need not be representative of $\sigma_M(x)$. At any rate, some sort of modification would surely be necessary if one were to apply this method to Ca and Sr since the pronounced Ramsauer minimum in these elements is shaped in large measure by the s-wave. In the elements considered here, the dip is quite small and will only affect the transport coefficients at very low values of E/N.

**Figure captions**

Fig. 1. Black curve: Elastic $\sigma_T(x)$ for sodium from equation (5). Black points: experiment of Kasdan *et al.* [23]. Red curve: $\sigma_M(x)$ from equation (5) with $\sigma'_0=1160$, $B'=40$. Red points: $\sigma_M(x)$ of Nakamura and Lucas [30].

Fig. 2. Calculated electron swarm drift velocities in sodium at $T=803°$ K fitted to two sets of experimental data of Nakamura and Lucas [29].

Fig. 3. Black curves: electron swarm mean energies in sodium at $T=803°$ K calculated from the Boltzmann equation for the two sets of data of figure 2. Red curves: corresponding mean energies calculated from the momentum-transfer approximation.

Fig. 4. Comparison of $\sigma_M(x)$ for sodium given by equation (5) with $\sigma'_0=1000$, to the function $\sigma_M(\varepsilon_m)$ calculated from equation (13).

Fig. 5. Elastic $\sigma_T(x)$ for magnesium from equation (14).

Fig. 6. Transmitted current of electrons through magnesium vapor. The black curve is from the experiment of Burrow *et al.* The red curve is obtained with the cross section of figure 5.

Fig. 7. The negative of the derivative of the $e^-$-Hg cross section of equations (16,18) (red curve) compared to the corresponding experimental recording of Johnston and Burrow [17,41].

Fig. 8. Upper curves: comparison of present $\sigma_T(x)$ defined by equations (16,18) and $\sigma_0=43.2$, $B=300$, with the experimental total cross section of Jost and Ohnemus [35]. Lower curves: comparison of $\sigma_M(x)$ defined by equations (16,18) to the momentum transfer cross section derived from swarm data by England and Elford [37].

Fig. 9. Calculated electron swarm drift velocity in mercury at $T=573°$ K fitted to the experimental data of England and Elford [37].

Fig. 10. Comparison of electron swarm mean energies in mercury at T=573° calculated from the Boltzmann equation (black curve) and from the momentum-transfer approximation (red curve).

Fig. 11. "Elastic" electron energy distribution at E/N=4, 4.7 and 5.7 Td, showing the formation and evolution of a second hump at higher energy.

Fig. 12. Comparison of $\sigma_M(x)$ for mercury given by equations (15,19) to the function $\sigma_M(\varepsilon_m)$ calculated from equation (13).

Fig.13(a) Three factitious "mercury" cross sections. The red and black profiles give rise to negative differential conductivity.

Fig.13(b) Electron swarm drift velocity w(E/N) corresponding to the cross sections of figure 13(a).

Fig. 14. Multi-peaked generalized Fano cross section similar to the theoretical $\sigma_T(x)$ for cesium of Bahrim and Thumm [11]. The red curve is the corresponding $\sigma_M(\varepsilon_m)$ calculated at T=5.8° K.